\documentclass[aps,twocolumn,showpacs,superscriptaddress]{revtex4}

\usepackage{amsmath,amssymb}
\usepackage{url}
\usepackage{graphicx,color}
\usepackage{tabularx}

\newcommand{\ket}[1]{\left| #1 \right\rangle}
\newcommand{\bra}[1]{\left\langle #1 \right|}
\newcommand{\etal}{\mbox{\emph{et.~al.}} } 

\graphicspath{{./figures/}}

\begin{document}

\title{Is spin-charge separation observable in a transport experiment?}
\author{T.~Ulbricht}

\affiliation{                    
  Institut f\"ur Theorie der Kondensierten Materie
  - Universit\"at Karlsruhe, 76128 Karlsruhe, Germany
}

\author{P.~Schmitteckert}

\affiliation{                    
   Institut f\"ur Nanotechnologie - Forschungszentrum Karlsruhe, 76021 Karlsruhe, Germany
}

\begin{abstract}
   We consider a one-dimensional chain consisting of an interacting
   area coupled to non-interacting leads.  Within the area, interaction
   is mediated by a local on-site repulsion.  Using real time evolution
   within the Density Matrix Renormalisation Group (DMRG) scheme, we
   study the dynamics of wave packets in this two-terminal transport
   setup.  In contrast to previous works, where excitations were
   created by adding potentials to the Hamiltonian, we explicitly
   create left moving single particle excitations in the right lead as
   the starting condition.
   Our simulations show that such a transport setup allows for a clear
   detection of spin-charge separation using time-resolved
   spin-polarised density measurements.
\end{abstract}
\pacs{71.10.Pm,71.10.Fd,72.15.Nj}

\maketitle

\section*{Introduction}
Although spin-charge separation (SCS) is undisputed to happen in
one-dimensional systems, its direct experimental observation and
verification is a subject of many works over the past years.  In one
dimension, the Fermi liquid picture of non-interacting quasi-fermions
is not appropriate \cite{Voit_RPP1995}.  Instead, in the
paradigmatic Tomonaga-Luttinger model and in the prototypical Hubbard
model collective bosonic spin and charge excitations are the
low-energy excitations.  These so-called spinons carry spin but no
charge and so-called holons carry the charge but no spin and spinons
and holons may propagate with independent velocities.

Up to now few experimental setups barely show direct evidence of
separate spin and charge excitations \cite{Halperin_JAP2007}. Beside
tunnelling experiments e.g. momentum-conserved tunnelling on
cleaved-edge overgrown GaAs/AlGaAs heterostructures
\cite{Auslaender_Steinberg_Yacoby_Tserkovnyak_Halperin_Baldwin_Pfeiffer_West_S2005}
and single-point tunnelling into carbon nanotubes
\cite{Tombros_vanderMolen_vanWees_PRB2006,Bockrath_Cobden_Lu_Rinzler_Smalley_Balents_McEuen_N1999},
there is angle-resolved photo emission spectroscopy (ARPES) on
quasi-one-dimensional SrCuO$_{2}$ material
\cite{Kim_Koh_Rotenberg_Oh_Eisaki_Motoyama_Uchida_Tohyama_Maekawa_Shen_Kim_NPhys2006}
and on the organic conductor TTF-TCNQ
\cite{Claessen_Sing_Schwingenschlogl_Blaha_Dressel_Jacobsen_PRL2002,Sing_Schwingenschlogl_Claessen_Blaha_Carmelo_Martelo_Sacramento_Dressel_Jacobsen_PRB2003}
which seem to show distinct spinon and holon branches in the spectral
function. Numerical simulations explaining the latter experiments were
done by Benthien \etal \cite{Benthien_Gebhard_Jeckelmann_PRL2004}. 

Here, we propose the following transport experiment: A
one-dimensional interacting region is connected to non-interacting
leads on both sides. In one of the leads a single electron with
average momentum $k_0$ is injected into the system. The excitation, by
passing through the interacting region, ends up in the other
non-interacting lead, where a spin-resolved and time-resolved
measurement of charge density is carried out.  This setup poses the
following questions which we answer in this paper: If one injects an
electron with definite momentum at some time $t_0$ into an interacting
system, where the separation of spin and charge is known to happen,
what happens to the electronic excitation within the interacting
region? And what kind of excitation emerges from the interacting area
at a later time? Will we see distinct excitations in spin and charge
density,  will we find spin and charge density to be recombined to
one full electronic excitation, or will we obtain an incoherent superposition
of many excitations?

Safi and Schulz \cite{Safi_Schulz_PRB1999} analyse such a transport
setup analytically for a spinless Tomonaga-Luttinger liquid model. For
the spinful case, however, they have to neglect ``umklapp processes
and the backscattering of electrons with opposite spin'' to sketch a
qualitative picture, thus looking at an idealised system, where spin
and charge currents are conserved separately. The first numerical
simulations for spin-charge separation were performed by Jagla \etal
\cite{Jagla_Hallberg_Balseiro_PRB1993} using exact diagonalisation for
16 sites.  Zacher \etal
\cite{Zacher_Arrigoni_Hanke_Schrieffer_PRB1998} used Quantum Monte
Carlo simulations to study the spin and charge susceptibilities of the
Hubbard model. Kollath \etal \cite{Kollath_Schollwock_Zwerger_PRL2005}
presented the first spin-charge separation calculations in the
framework of adaptive time dependent DMRG (td-DMRG). This calculation
was redone by Schmitteckert \cite{Schmitteckert_HPC2007}, keeping up
to 5000 states per DMRG block showing that the adaptive time evolution
scheme is prone to large numerical errors in the long time limit.  In
the same line, Schmitteckert and Schneider
\cite{Schmitteckert_Schneider_HPC2006} used over 10000 states per DMRG
block to show SCS in a 33 site system with periodic boundary
conditions which avoids Friedel oscillations.

In order to avoid the numerical problem
associated with the adaptive time evolution scheme and to avoid the
numerical costs of a pure full td-DMRG we apply a hybrid of both
methods, see \cite{Schmitteckert_HPC2007,Boulat_Saleur_Schmitteckert_PRL2008}.  
We first use the standard ground state DMRG
\cite{White_PRL1992} to obtain our initial state. We then perform
typically ten time steps using the full td-DMRG
\cite{Schmitteckert_PRB2004} and continue with an adaptive td-DMRG
\cite{Vidal_PRL2004,White_Feiguin_PRL2004,Daley_Kollath_Schollwock_Vidal_JSMTE2004}
scheme. For recent reviews on DMRG,
see \cite{DMRG_reviews}.
In addition, two main objectives distinguish our work from others that
use real-time analysis and model strongly interacting fermionic
systems.  1)~Previous works either treated only homogeneous
interacting systems
\cite{Jagla_Hallberg_Balseiro_PRB1993,Schmitteckert_PRB2004,
  Kollath_Schollwock_Zwerger_PRL2005} or modelled optical traps
\cite{Kollath_Schollwock_Zwerger_PRL2005}.  In contrast to that our
system is a two-terminal setup modelling a transport experiment.
2)~While in earlier works
\cite{Schmitteckert_PRB2004,Kollath_Schollwock_Zwerger_PRL2005}
excitations were studied based on the modification of local
potentials, we create an excited state by explicitly adding one
electron to the system and we monitor the time evolution of such an
excited state.

\paragraph*{Outline}
First we motivate our creation of an excitation and outline how the
numerical method is used to measure the observables. This is
accompanied by bulk simulations on a non-interacting and an
interacting system.  Then we apply the method to the transport setup.

\section*{Motivation and method}
\begin{figure}[h]
  \centering
  \begin{tabularx}{\linewidth}{>{\hsize=0.5\hsize}X>{\hsize=0.5\hsize}X}
    \includegraphics[width=0.5\linewidth]{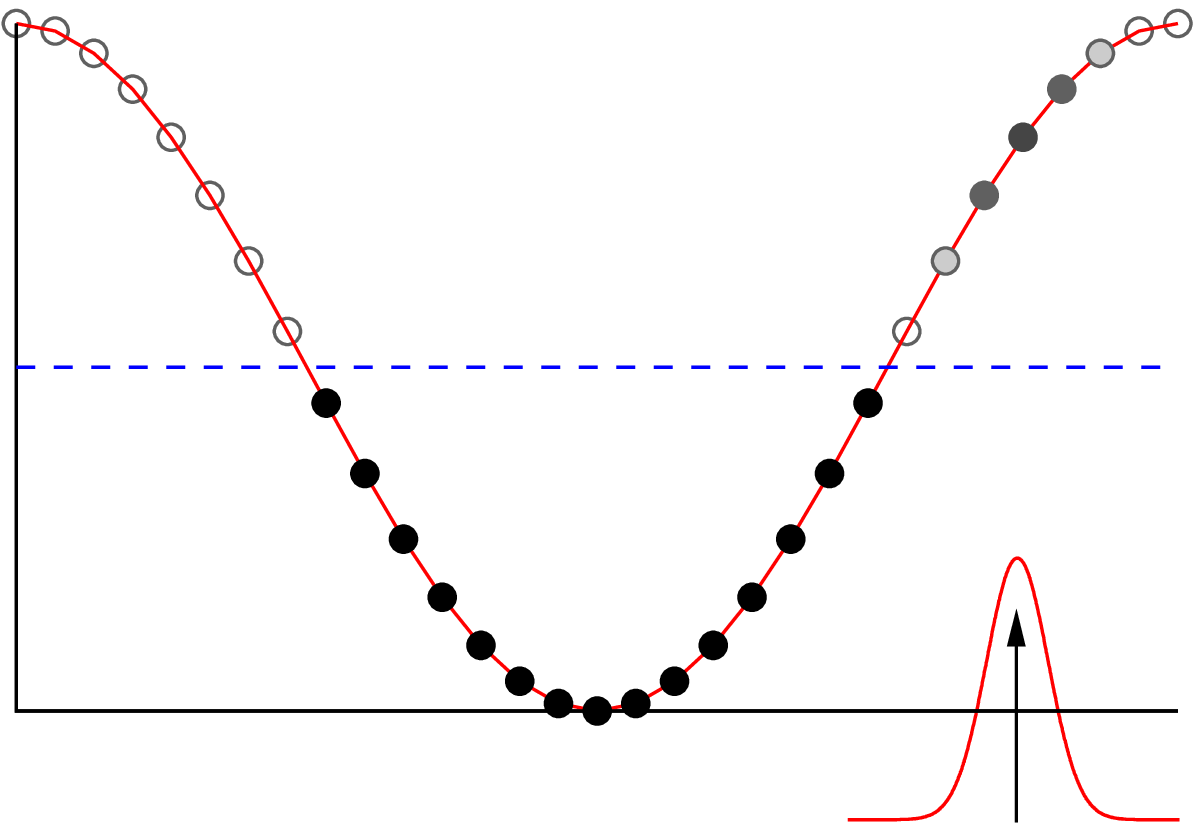}
    &
    \includegraphics[width=0.5\linewidth]{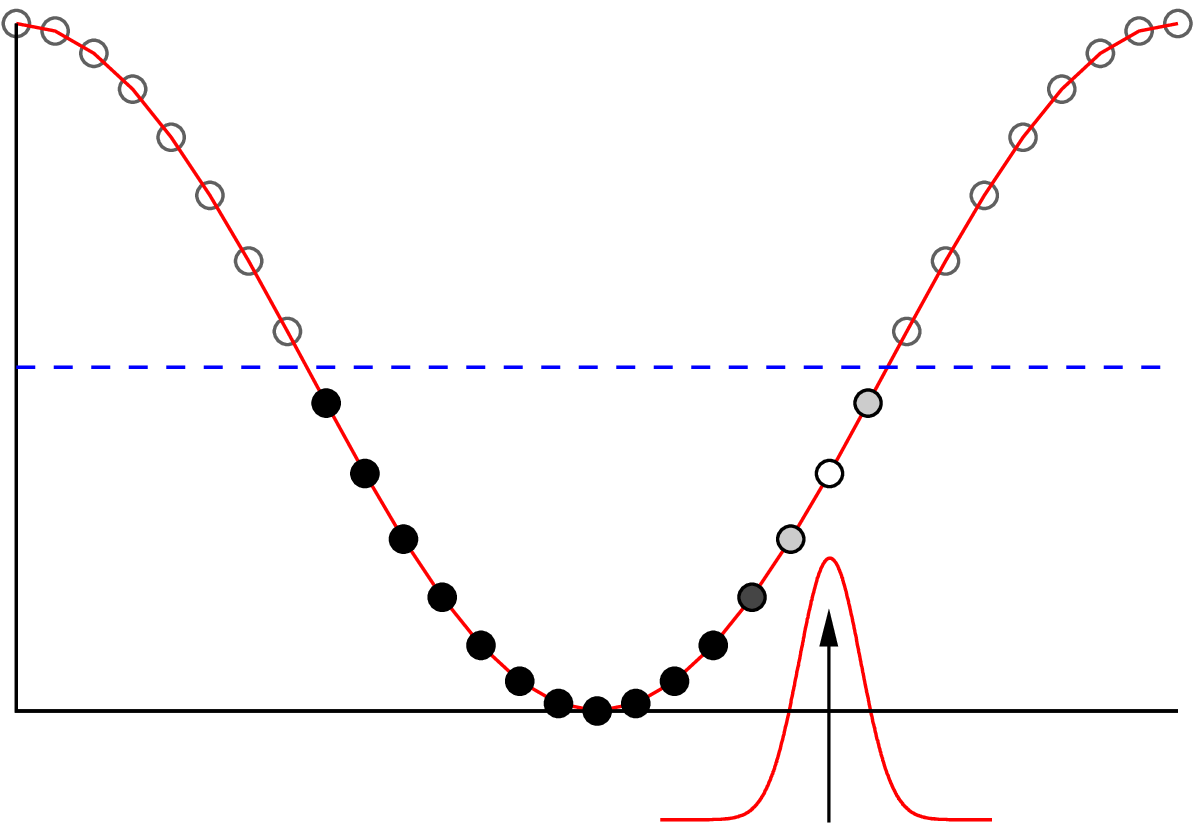}
   \end{tabularx}
   \caption{(Colour online) Visualisation of the creation of an
     electron (left) or a hole (right) by indicating the occupation of
     single particle levels in the dispersion relation of the
     non-interacting problem. The dotted line represents the Fermi energy.}
  \label{fig:gaussian}
\end{figure}

An excitation, created at a given site $x_0$ in the tight-binding
model by using $c_{x_0}^{\dagger}$, excites all eigenmodes in the
diagonal basis of momentum eigenstates of the tight-binding model.
This is not particularly useful since we want to consider electrons
with a definite momentum.  On the other hand, creating an excitation
using $c_{k_0}^{\dagger}$ local in the momentum space is again an
eigenstate of the system and invariant under time evolution.
Therefore in the ground state $\ket{\Psi}$ of a half-filled
tight-binding model (for each spin specie $\sigma = \pm \frac{1}{2}$),
\begin{equation}
  \label{eq:tightbindingbulk}
  H_{\text{tb},\sigma} = - \sum_{x} \left( c_{x-1,\sigma}^{\dagger}
    c^{}_{x,\sigma} + \text{h.c.} \right)
\end{equation}
we create an excitation with spin $\sigma$ by using a Gaussian
distribution of creation (annihilation) operators, with momenta
centred around $k_0>k_F (k_0<k_F)$:
\begin{align}
  g^{(\dagger)}_{\sigma}(k_{0}) &= \sum_{k}
  {\mathrm e}^{-\frac{(k-k_{0})^{2}}{2\sigma^2_{0}}} c^{(\dagger)}_{k,\sigma} \\
  \label{eq:gaussianexcitation}
  &= \sum_{x=1}^{M} \left( {\mathrm e}^{(-2\sigma^2_{0}\pi^2 x^{2})}
  \right) \left( {\mathrm e}^{\frac{2\pi ix}{M}k_{0}}
  \right)c^{(\dagger)}_{x,\sigma}
\end{align}
where $k$ are the momentum eigenstates $k \in \{-\pi+
\frac{2\pi}{M},\dots,\pi\}$, and $\sigma_{0}$ is the width of the
excitation in momentum space. Additionally, the excitation will be
centred around $x_0$ and normalised such that integrating over all
$k$ will resolve to exactly one added electron (hole). A visualisation
of this construction is shown in figure \ref{fig:gaussian}.  An
excitation created in this way
\begin{align*}
  \ket{\Psi^{+1}} =
  \frac{1}{\sqrt{C}}g_{\sigma}^{(\dagger)}(k_{0}) \ket{\Psi} 
\end{align*}
with $C = \bra{\Psi} (g_{\sigma}^{(\dagger)}(k_{0}))^{\dagger}
g_{\sigma}^{(\dagger)}(k_{0}) \ket{\Psi}$ will move through the system
in one direction with a group velocity $v(k_{0}) = \frac{\partial
  \epsilon(k)}{\partial k}|_{k_{0}}$.  Using the Schr\"odinger picture
and evolving our initial state $\ket{\Psi^{+1}(t)} = {\mathrm e}^{iH
  t}\ket{\Psi^{+1}}$, the measured observable is the spin-resolved
density 
\begin{equation}
n_{\sigma}(x,t) = \bra{\Psi^{+1}(t)} c_{x,\sigma}^{\dagger}
c^{}_{x,\sigma} \ket{\Psi^{+1}(t)}
\end{equation}
which allows the measurement of charge density $n_{c} = n_{\uparrow} +
n_{\downarrow}$ and spin density $ n_{s} = \frac{1}{2}(n_{\uparrow} -
n_{\downarrow} )$ at any time $t$ as a function of $x$.

In the following figures, we use DMRG \cite{White_PRL1992} to generate
the ground state $\ket{\Psi}$ and the excited state $\ket{\Psi^{+1}}$.
For both, the full and the adaptive td-DMRG, the time evolution is
performed by a Krylov subspace based evaluation of the time evolution
$ \ket{\Psi^{+1}(t+\Delta t)} = {\mathrm e}^{iH\Delta
  t}\ket{\Psi^{+1}(t)}$, for details see
\cite{Schmitteckert_PRB2004,Schmitteckert_HPC2007,Boulat_Saleur_Schmitteckert_PRL2008}.
The time steps were $\Delta t = 0.5$ (in units of time) for both
methods.

We use hard-wall boundary conditions in all simulations and all
energies are in units of the hopping parameter which is set to 1 in
\eqref{eq:tightbindingbulk}.  We also calculate the static ground
state density $n_{0,\sigma}(x) =
\bra{\Psi}c_{x,\sigma}^{\dagger}c^{}_{x,\sigma}\ket{\Psi}$ at time
$t=0$ and form the corresponding charge and spin densities $n_{0,c},
n_{0,s}$. Then we plot the quantities $n_{s}(x,t) - n_{0,s}(x)$ and
$n_{c}(x,t) - n_{0,c}(x)$. This trick evens out all stationary
oscillations already present in the ground state, like the Friedel
oscillations resulting from using hard-wall boundary conditions.
Additional $2k_F$-oscillations resulting from the density distortion
of the added excitation are evened out using a three-point average.

\begin{figure}[ht]
  \centering
    \includegraphics[width=\linewidth]{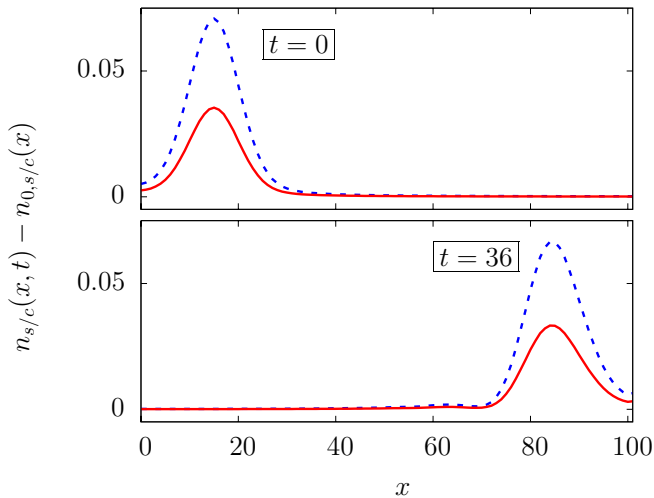}
    \caption{(Colour online) Spin (solid,red) and charge (dashed, blue)
      densities of a td-DMRG simulation of one additional up-electron
      over the background of a half-filled tight-binding model for
      each spin specie. The Gaussian excitation on the left at time
      $t=0$ (upper diagram) moving to the right at $t=36$ (lower
      diagram) retains its shape. Averaging is described in the text.}
  \label{fig:non-int}
\end{figure}

In our non-interacting (tight-binding) reference system
\eqref{eq:tightbindingbulk} in figure \ref{fig:non-int} we create an
electron at the real space position $x_0 = 15$ (upper diagram) which
travels to the right (lower diagram).  The system of size $M=102$ is
at half-filling plus one electronic excitation $N_{\uparrow}=51+1$.
The  $N_{\downarrow}=51$ down-spins are, of course, irrelevant in the
non-interacting case and constant in the time evolution. The
excitation's momentum is centred around $k_{0} = \pi/2 + 0.1 $ with
width $\sigma_0 = 0.03$. With this width, we ensure that the momentum
distribution is far away (compared to the width) from the Fermi
surface $k_{0} \sim k_{F} + 3\sigma_0$ but still as close as possible
to the linear regime of the cosine band at $k_F$.
The snapshots at $t=0$ and $t=36$ in figure \ref{fig:non-int} of spin
and charge density serve as a proof-of-concept for the numerics. They
show the expected synchronous motion of the electronic excitation with
group velocity $v_G = 1.93\pm0.02$. A delta-function-shaped excitation
in the thermodynamic limit of this system would have a group velocity
of $v_G(k_0)\sim1.99$. Since we are in a finite system, the
discretisation and the cut-offs at the Fermi energy and the band edge
directly influence the actual shape of the initial density
distribution. Also, the non-linearity of the band affects the packet
propagation via the width and position of the momentum
distribution. The (otherwise perfect time invariant) Gaussian shape is
distorted and the packet is slowed down.
This system was calculated using $N_{\text{cut}}=1000$ states per DMRG
block, applying 10 time steps in the full td-DMRG and keeping $1000$
states for the consecutive adaptive td-DMRG. The discarded entropy
never exceeds the order of $10^{-5}$. Comparing to an exact
diagonalisation (ED) we define the peak relative error in the densities by
\begin{equation}
  |\delta n_p(t)| \equiv \max_{x} \left( \frac{n_{\text{ED}}(x,t) -  
      n_{\text{DMRG}}(x,t)}{n_{\text{DMRG}}(x,t)}\right)
\end{equation}
This quantity reaches from initial $ |\delta n_{p} (0)| = 7 \cdot
10^{-5}$ peak relative error to $|\delta n_p (36)| = 7 \cdot 10^{-3}$
peak relative error at $t=36$.

\begin{figure}[h]
  \centering
    \includegraphics[width=\linewidth]{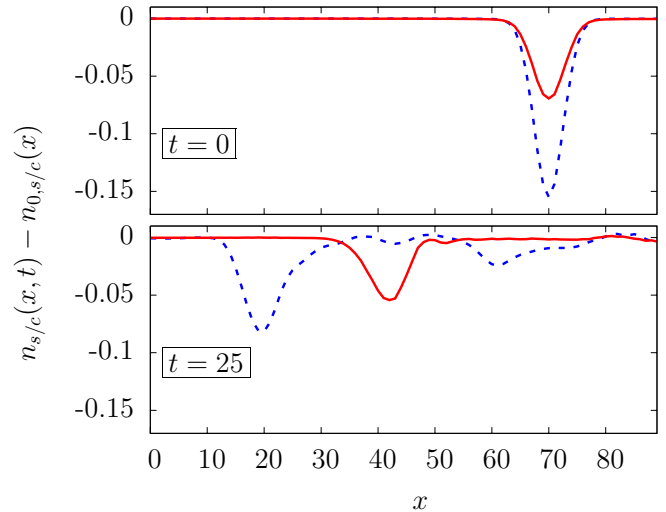}
    \caption{(Colour online) Spin and charge density (see legend of
      fig.  \ref{fig:non-int}) in a Hubbard model at time $t=0$ (upper
      diagram) and $t=25$ (lower diagram) show spin-charge
      separation. This is data from a td-DMRG simulation on 90 sites
      system with $N_{\uparrow}=41-1$, $N_{\downarrow}=41$ electrons
      and a Hubbard interaction of $U=4$.}
  \label{fig:int}
\end{figure}

Now we repeat the simulation in a Hubbard model bulk system
\begin{equation}
  \label{eq:hubbardbulk}
  H_{\text{bulk}} =
  \sum_{\Sigma=\uparrow,\downarrow}H_{\text{tb},\sigma}  +  U \sum_x
  \left( n_{x,\uparrow} -  \frac{1}{2} \right)\left( n_{x,\downarrow} 
    - \frac{1}{2} \right).
\end{equation}
While the previous excitation was a superposition of well defined
eigenstates in the tight-binding model, it is not clear which
eigenmodes are excited by adding/removing an electron from an
interacting system. Nevertheless, using the same $c_{\sigma
  x}^{(\dagger)}$ in equation \ref{eq:gaussianexcitation} to create an
excitation still results in a Gaussian shape in spin and charge
densities (see figure \ref{fig:int}), although the corresponding
Fourier transformed operators $c_{\sigma k}^{(\dagger)}$ no longer
create single particle eigenstates of the system.
This time we insert an excitation corresponding to one (up-spin) hole
into the ground state of a repulsive particle-hole symmetric Hubbard
model \eqref{eq:hubbardbulk} with $U=4$ at a filling of
$\frac{41}{90}\approx 0.46$ with a central momentum $k_{0} = \pi/2 -
0.5 \sim k_{F} - 7\sigma_0 $ and a width of $\sigma_{0} \sim
0.053$. The calculations were carried out with $N_{\text{cut}}=2500$ states
kept per DMRG block in the full  and adaptive  time dependent DMRG.
One clearly observes SCS driven by the Hubbard interaction between
$t=0$ (upper diagram) and $t=25$ (lower diagram). We extract the group
velocities from the main peaks to be $v_S=1.12\pm 0.05$ and
$v_C=2.03\pm 0.05$. The velocity is constant through out the
simulation and the error stems from the read-off error. The one
dimensional Hubbard model can be solved exactly \cite{Lieb_Wu_PRL1968}.
From the analytic expression by Coll \cite{Coll_PRB1974}, the
spin and charge velocity of the low energy spinons and holons can be
derived as done, e.g. by Schulz \cite{Schulz_IJMPB1993}.
The excitation's momentum of $k_0$ corresponds to a band filling of
$n=0.68$ in units of the hopping. Given the error estimate above, it
is accurate enough to read off a charge velocity of $v_{C,\text{one
    holon}} \sim 2.05$ and $v_{S,\text{one spinon}} \sim 1.15$
(cf. figure 11 of the first paper of Ref.  \cite{Schulz_IJMPB1993}).
Although we are with 90 sites far from simulating the thermodynamic
limit, the group velocities of our spin and charge density fit
remarkably well to the Bethe Ansatz results for the velocity of one
spinon and one holon. A similar good agreement was found for a
spinless Luttinger liquid simulation \cite{Schmitteckert_PRB2004}.

Our electronic excitation with a definite momentum $k_{0}$ created a
unidirectional running peak in the non-interacting system. Here,
however, our initial excitation creates both a right and a left moving
density peak in the charge sector. At time $t=25$, the right moving
charge peak has already been reflected on the right wall and is
located at around $x=60$. The spin density, in contrast, displays
predominantely a left moving peak. Finally, there is charge density
accumulation trailing the main charge peak and at the location of the
spin peak. These might be more complicated features of the interaction
or result from the finiteness of the system as in the non-interacting
example.

Following the previous line of arguments, it seems reasonable to do a
transport setup that uses non-interacting leads as a starting point.
We let this excitation travel through an interacting region where SCS
occurs and then let the resulting excitations travel again into a
non-interacting area. In the starting lead our electronic excitation
is well understood. When this excitation travels into the interacting
region, we know the momentum and energy of the injected wave packet
(one electron or one hole) but the questions arise: what kind of
excitation leaves the structure and what can be measured on the
opposite (non-interacting) lead?
\begin{figure}[h]
  \centering
    \includegraphics[width=\linewidth]{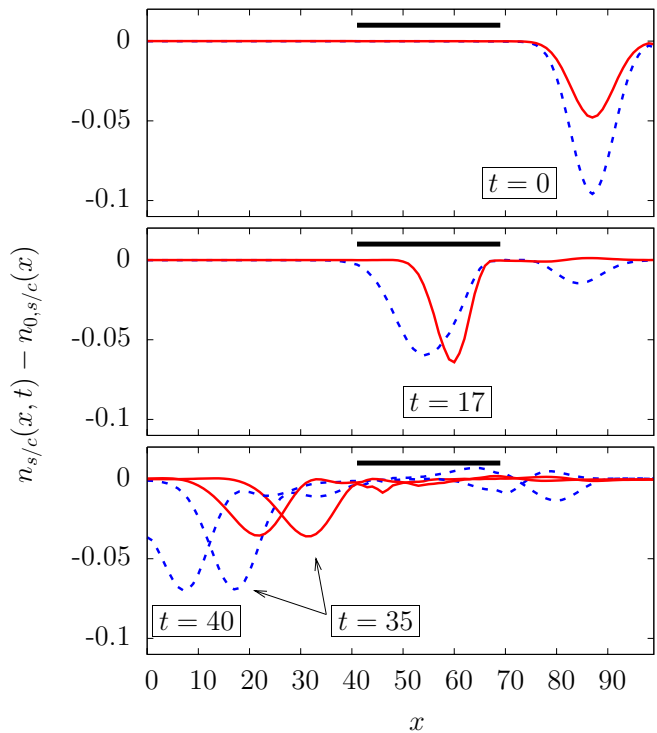}
    \caption{(Colour online) Excitation in a transport experiment: a
      created hole in the right lead (upper diagram, at $t=0$) passes
      the interacting nanostructure (black bar) undergoing SCS
      (central diagram, at $t=17$). At $t=35,40$ (lower diagram) spin
      and charge densities travel independently with equal velocity in
      the left lead.}
  \label{fig:transport}
\end{figure}

We use a system of $M=100$ sites, split into 41,
29 and 30 sites for the different areas and a total of
$N_{\uparrow}=48-1$, $N_{\downarrow}=48$ electrons. 
The corresponding Hamiltonian reads
\begin{multline}
  \label{eq:transport}
  H_{\text{transport}} =
  \sum_{\sigma=\uparrow,\downarrow}H_{\text{tb},\sigma} + V_{\mathrm{gate}}
  \sum_{x=42..70} (n_{x,\uparrow} + n_{x,\downarrow} ) \\
  + U \sum_{x=42..70} \left( n_{x,\uparrow} - \frac{1}{2}
  \right)\left( n_{x,\downarrow} - \frac{1}{2} \right),
\end{multline}
where $H_{\text{tb},\sigma}$ was defined in \eqref{eq:tightbindingbulk} and $U=2.5$.
A gate voltage on
the interacting area levels the Fermi surface to a half-integer number
of up and down electrons on the structure
($N_{\uparrow}=N_{\downarrow}=12.5$) and minimises reflections into
and out of the area (see figure \ref{fig:transport}). This local
chemical potential $V_{\mathrm{gate}}$ in \eqref{eq:transport} was pre-calculated in a self-consistent way. Thus we
have a Hubbard model at $\sim 0.43$-filling in the central (black bar)
area and half-filled tight-binding leads.  We also chose parameters
$k_{0} = 0.43\pi - 2\sigma_{0}$, $\sigma_{0} = 0.03$ to maximise
tunnelling while staying as close as possible to a quasi-linear
dispersion. The number of states kept was $N_{\text{cut}} = 2500$.
Figure \ref{fig:transport} shows the time evolution of spin and charge
densities for several times. There are reflections of charge density
at both edges of the structure. They are suppressed by the choice of
the chemical potential as discussed above.  The majority of spin and
charge density transmits through the nanostructure.  SCS clearly
happens inside the interacting region (central diagram at $t=17$) and
SCS remains intact after leaving the interacting area into the
opposite lead (lower diagram at $t=35$ and $t=40$).  Here spin and
charge now travel along separately but again with identical constant
velocity.
\section*{Conclusions}
In our simulation of a microscopic experiment on an interacting
nanostructure we saw that the spin and charge excitation of an
injected electron can be separated as it is known to happen in
one-dimensional interacting systems.  The spin and charge separation
can be directly observed by a time-resolved measurement of the
spin-polarised density. There is no need for spin and charge to
recombine to a complete electron (hole) for a measurement of a single
electron (hole).  Note that the charge and spin excitations in the
non-interacting leads are valid excitations of the Fermi liquid. They
are not \emph{elementary} excitations of the Fermi liquid, instead
they are a composition of (at least) two excitations of the Fermi
liquid. As an example, think of the combinations
$(c^{\dagger}_{k,\uparrow} \pm c^{\dagger}_{k,\downarrow})$. However
the resulting distribution of separated densities originated from a
single particle injection into the Fermi sea.  Thereby, we confirm for
the Hubbard model, what Safi and Schulz \cite{Safi_Schulz_PRB1999}
sketched for a low-energy Luttinger liquid transport setup with
restricted interaction. It is interesting that our small
  system already shows the fingerprint of Luttinger Liquid description
  expected in the thermodynamic limit. Also note that even for highly
energetic incoming electrons we find the spin-charge separation,
albeit obscured by band effects.  With the td-DMRG method even more
realistic models, like the extended Hubbard model with a finite
next-nearest neighbour interaction could be investigated.

Experimentally, the measurement of the densities would have to happen
on a time scale which is set by the Fermi velocity and the length of
the interacting system. For nanoscaled systems and metallic excitation
velocities tens of Terahertz resolution would be required, which might
be achieved using optical techniques. Lowering the Fermi, holon and
spinon velocities would enhance the detection possibilities.
It is of general
interest to know, whether a series of quantum dots or any other
strongly correlated system with a much smaller Fermi velocity can
provide the ingredients for the realisation of such a transport setup.

\begin{acknowledgments}
  We thank \mbox{Dmitry} \mbox{Aristov}, \mbox{Rafael} \mbox{Molina},
  \mbox{Ronny} \mbox{Thomale}, and \mbox{Peter} \mbox{W\"olfle} for
  valuable discussions and we acknowledge the support by the Center
  for Functional Nanostructures (CFN), project B2.10.  The DMRG
  calculations have been performed on HP XC4000 at Steinbuch Center
  for Computing (SCC) Karlsruhe under project RT-DMRG.
\end{acknowledgments}

\end{document}